\documentclass[12pt]{iopart}
\usepackage{bm}
\usepackage{caption}
\usepackage{subcaption}
\expandafter\let\csname equation*\endcsname\relax
\expandafter\let\csname endequation*\endcsname\relax
\usepackage{amsmath}
\usepackage{mathtools}
\begin{document}

\title[]{\boldmath Inflation Alternative via \\the Gravitational Field of a Singularity}

\author{Michael Zlotnikov}

\address{Department of Physics, Center for Theoretical Physics,
Columbia University, 538 West 120th Street, New York, NY 10027, USA}
\ead{michael\_zlotnikov@alumni.brown.edu}
\vspace{10pt}

\begin{abstract}
We explore the scenario that the observable universe emerged from the vicinity of a negative mass ring singularity, and all content of the universe travels at the same group velocity close to the speed of light on a geodesic trajectory along the axis of rotation of the singularity. In appropriate coordinate parametrization and evaluated on the trajectory, we find that the metric tensor in the vicinity of the trajectory exhibits a conformal scale factor $\bm{a}(\eta)$ with contraction and subsequent expansion properties that solve the horizon problem. We then introduce a static flow of gravitating radiation along the trajectory (perturbatively with respect to the mass scale of the singularity) to model a homogeneous radiation dominated universe. Solving the Einstein field equations with a physically motivated ansatz of metric perturbation then reveals that the effective conformal scale factor indeed grows asymptotically with the same power law as expected in a conventional radiation dominated universe.
\end{abstract}



\section{Introduction}
\label{sec:intro}

The horizon problem \cite{Rindler:1956yx} is concerned with the fact that different regions of the Cosmic Microwave Background (CMB) \cite{Dicke:1946yx,Dicke:1965zz}, emitted at the surface of last scattering, possess surprisingly uniform temperature. Assuming a generic cosmic evolution, these regions could not have been in causal contact with each other, and yet their almost equal temperatures suggest that they must have thermalized at some point in the past. A possible solution to this problem is the inflationary universe paradigm \cite{Starobinsky:1980te,Guth:1980zm}, in which the universe undergoes a period of accelerated expansion at early times, taking space-time regions that previously interacted with each other out of causal contact, which then re-enter the cosmological horizon upon later cosmic evolution. The process of inflation is typically assumed to be driven by a space-filling inflaton field (see for example \cite{Guth:1980zm,Kazanas:1980tx,Sato:1980yn,Linde:1981mu,Albrecht:1982wi,Linde:1983gd,Mukhanov:1981xt,Boubekeur:2005zm,Dvali:2001fw,Adams:1992bn,Silverstein:2008sg,Bezrukov:2007ep,Murayama:1992ua,Ellis:2013xoa}). At the end of inflation all other fields are thinned out and the universe is re-populated with standard model constituents through the decay process of the inflaton field called reheating \cite{Kofman:1997yn,Bassett:2005xm}.

In this work we explore an alternative way to resolve the horizon problem by assuming that the observable universe has been emitted from the vicinity of a negative mass ring singularity in a Kerr-like solution of the Einstein field equations,\footnote{Negative mass regions are commonplace in exact solutions of General Relativity, and can be argued to be physically meaningful. See e.g.\ \cite{Chardin:1996qs} for a motivation.} and all content of the universe travels at approximately the same group velocity close to the speed of light along a radial geodesic parallel to the Kerr spin axis. The relative motion with respect to the singularity manifests itself locally on the geodesic in terms of an effective cosmological scale factor with specific contraction and expansion properties solving the horizon problem. 

Subsequently, introducing a static flow of gravitating radiation, with energy density parametrically smaller than the mass of the singularity, we show that a leading order perturbative correction to the metric tensor leads to the same asymptotic late conformal time scaling of the effective scale factor as expected in a  Friedmann–Lemaître–Robertson–Walker (FLRW) radiation dominated universe. The transition to a matter dominated universe could then be made in a conventional formalism, without invoking a singularity in the far past.

As described by Poisson and Israel \cite{Poisson:1990eh}, the inner region of a Kerr black hole may develop prohibitive properties when radiation is present, such as mass inflation at a Cauchy horizon. However, we emphasize that this work is concerned with the vicinity of a geodesic passing through a region Poisson and Israel call a \textit{future asymptotically flat universe}, far away from the inner region of the black hole. For completeness, we may nevertheless feel compelled to think of possible issues in the inner black hole region. In that case, the reader is invited to consider only an ingoing flux of radiation entering the black hole from the viewpoint of the \textit{past asymptotically flat universe}, with a negligible outgoing flux. In that scenario, the infinitely blueshifted Cauchy horizon merges with the infinitely redshifted inner horizon, canceling out and preventing any mass inflation effect from occurring.

While in the inflation paradigm the entire universe starts out as a small blob of vacuum energy, in this work the observable universe starts out as a small region within a seemingly implausible much larger preexisting universe containing a singularity that appears to be naked and may seem unphysical. However, by the cosmic censorship principle a singularity is only unphysical if it can actually be directly observed. Note that the Kerr spacetime is static, which means that the singularity is not radiating off energy, and therefore there is no signal emerging from the singularity to be detected by any local observer. In fact, it is impossible for the space close to the negative mass singularity in a \textit{static} spacetime to be anything but completely empty due to the singularities' gravitational repulsion. The static stream of radiation discussed in this work arrives axially from infinity of the positive mass Kerr spacetime region, passes through the \textit{eye of the needle} of the ring singularity, emerges axially in the negative mass region and escapes towards infinity. This radiation that actually reaches the negative mass region (radiation that is of interest in section \ref{sec:radiation} below), by definition cannot carry any local information about the singularity since it successfully escaped it.

This paper is organized as follows:

In appendix \ref{sec:prelim} we recall the basic tools of General Relativity to be used in this work.

Section \ref{sec:empty} introduces an appropriate cylindrical change of coordinates, parametrizing the vicinity of a radial geodesic trajectory parallel to the axis of rotation in a negative mass Kerr space-time. Evaluating the metric on the geodesic trajectory and expressing the solution in conformal time reveals that at leading order in the vicinity of the trajectory the metric possesses an effective conformal scale factor $\bm{a}$. The scale factor $\bm{a}$ starts at unity, contracts to a parametrically small number, depending on the Kerr spin parameter and mass $a/M\to0$, and subsequently asymptotically expands back to unity in the case of an empty universe. The initial contraction and subsequent expansion of the scale factor explains how seemingly separated regions entering the cosmological horizon during the expansion phase were in causal contact previously.

In section \ref{sec:radiation} we perturbatively add a static flow of gravitating radiation traveling along the radial geodesic. With an appropriate perturbation ansatz for the metric tensor, we solve the Einstein equations to leading order in the vicinity of the radial trajectory and  small Kerr spin parameter $a/M\to0$. The resulting correction to the effective conformal scale factor turns out to be given by an integral over ratios of hypergeometric functions. At late conformal time $\eta\gg M$, the effective conformal scale factor then turns out to scale as $\sim\eta$, which is the expected scaling for a radiation dominated universe.

We conclude in section \ref{sec:conclusion}, and offer a few suggestions for further investigations that go beyond the scope of this paper.

\section{Empty Universe}
\label{sec:empty}
We start with the \textit{negative mass} region of the Kerr metric \cite{Kerr:1963ud} in Boyer-Lindquist coordinates, employing the parameter $M>0$:\footnote{Compared to the regular Kerr black hole metric, this differs by the sign in front of the mass parameter $M$, and is equally an exact solution of the vacuum Einstein field equations.}
\begin{align}
\label{kerr}
-d\tau^2=&-\frac{\Delta_r
   }{\Sigma_r}\left(dt-a d\phi \sin ^2(\theta )\right)^2+\Sigma_r \left(\frac{dr^2}{\Delta_r}+d\theta^2\right) +\frac{\sin ^2(\theta )}{\Sigma_r} \left(a dt-d\phi \left(a^2+r^2\right)\right)^2,
\end{align}
where
\begin{align}
\Delta_r=a^2+2 M r+r^2~~~,~~~\Sigma_r=a^2 \cos ^2(\theta )+r^2~~~\text{and}~~~a=\frac{J}{M},
\end{align}
with spin $J$ of the ring singularity. Note that we are using the geometrized units convention, setting the gravitational constant as well as the speed of light to unity $G=1,c=1$. Since there is no relative minus sign in the $\Delta_r$ factors, this space-time does not exhibit any black hole horizons. Therefore, particles can escape from the \textit{`eye of the needle`} of the ring singularity to infinity, which will be an important feature in the following.

We work towards understanding how the metric behaves locally, when evaluated on an outgoing radial geodesic trajectory and expressed in conformal coordinates at leading order in the vicinity of the trajectory. This shall facilitate a comparison with a conventional cosmological metric of an unperturbed universe. 

As a first step, we introduce deformed cylindrical coordinates, along with an expansion around small cylindrical radius, to conveniently describe the vicinity of the straight line trajectory. By \textit{deformed cylindrical coordinates} at small radius we mean substitutions:
\begin{align}
r\sin\theta &= \bm{r} f_1(z) + \bm{r}^3 f_3(z) + \mathcal{O}\left(\bm{r} ^5\right)\,,\\
r\cos\theta &= z + \bm{r}^2 f_2(z) + \mathcal{O}\left(\bm{r} ^4\right)\,,
\end{align}
where our deformation allows the $\bm{r}$ term on the right hand side to have a $z$-dependent coefficient, as well as includes $\bm{r}^2$ and $\bm{r}^3$ terms that are sub-leading for small $\bm{r}$. Note that $\bm{r}$ is related to the original Boyer-Lindquist $\theta$ angle variable, while $z$ is related to the original $r$ radius variable; however, in the following $\bm{r}$ and $z$ will serve as the radius and axis variables respectively. The functions $f_1(z),\,f_2(z),$ and $f_3(z)$ should be chosen such that this coordinate transformation is invertible in the small $\bm{r}$ region of interest. We make the following particular choices:
\begin{align}
f_1(z)=\frac{z}{\sqrt{q(z)}}~~~,~~~f_2(z)&=\frac{M}{2q(z)}~~~,~~~f_3(z)=\frac{a^4 (M+z)+a^2 z^2 (3 M+z)}{2 \left(a^2+z^2\right)^2
   q(z)^{3/2}}\,,\\
	q(z)&=a^2+z(2M+z)\,,
\end{align}
because they lead to a useful form of the resulting metric.
 In terms of a direct change of coordinates $(r, \theta)\to(\bm{r},z)$ for $z\geq 0$, and an expansion around small $\bm{r}\ll a^2/M$, we have:\footnote{Here, small cylindrical radius $\bm{r}$ corresponds to small $\theta$, while the original oblate spheroidal radius $r$ approaches $z$ in that limit.}
\begin{align}
\label{crdtrafo}
r=& \,z+\bm{r} ^2\frac{(M+z) }{2 \left(a^2+z (2 M+z)\right)}+\bm{r} ^4\frac{\left((2 M+3 z) a^4+2 z^2 (4 M+z) a^2-z^4 (2 M+z)\right) }{8
   \left(a^2+z^2\right)^2 \left(a^2+z (2 M+z)\right)^2}+\mathcal{O}\left(\bm{r} ^5\right),\notag\\
\theta =&\, \bm{r}\frac{1 }{\sqrt{a^2+z (2 M+z)}}+\bm{r} ^3\frac{a^4+(3
   M-z) z a^2-z^3 (3 M+2 z) }{6 \left(a^2+z^2\right)^2 \left(a^2+z (2 M+z)\right)^{3/2}}+\mathcal{O}\left(\bm{r} ^5\right),
\end{align}
resulting in the following differential relations
\begin{align}
dr&=\, dz+\bm{r}\frac{ M+z  }{a^2+z (2 M+z)}d\bm{r}-\bm{r}
   ^2\frac{-a^2+2 M^2+z^2+2 M z }{2 \left(a^2+z (2 M+z)\right)^2}dz\notag\\
	&\,+\bm{r} ^3\frac{(2 M+3 z) a^4+2 z^2 (4 M+z) a^2-z^4 (2 M+z) }{2
   \left(a^2+z^2\right)^2 \left(a^2+z (2 M+z)\right)^2}d\bm{r}+\mathcal{O}\left(\bm{r} ^4\right),\\
d\theta&= \frac{d\bm{r}}{\sqrt{a^2+z^2+2 M z}}-\frac{dz (M+z) \bm{r} }{\left(a^2+z (2
   M+z)\right)^{3/2}}+\frac{d\bm{r} \left(a^4+(3 M-z) z a^2-z^3 (3 M+2 z)\right) \bm{r} ^2}{2 \left(a^2+z^2\right)^2 \left(a^2+z (2
   M+z)\right)^{3/2}}\notag\\
	&-\bm{r} ^3\frac{  3 a^6+\left(M^2+12 z M+4 z^2\right) a^4+z^2 \left(12 M^2+6 z M-z^2\right) a^2-z^4 \left(5 M^2+6 z
   M+2 z^2\right) }{2 \left(a^2+z^2\right)^3 \left(a^2+z (2 M+z)\right)^{5/2}}z\, dz\notag\\
	&+\mathcal{O}\left(\bm{r} ^4\right).
\end{align}

\begin{figure}
\centering
\begin{subfigure}{.5\textwidth}
  \centering
  \includegraphics[width=\textwidth]{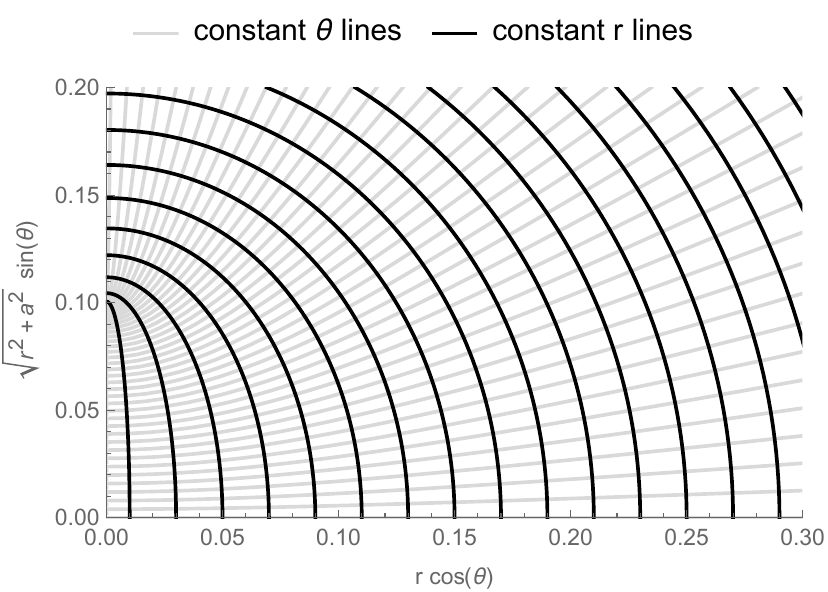}
  \caption{$\scriptstyle \text{Boyer-Lindquist oblate spheroidal coordinates}$}
  \label{figgs:sub1}
\end{subfigure}%
\begin{subfigure}{.5\textwidth}
  \centering
  \includegraphics[width=\textwidth]{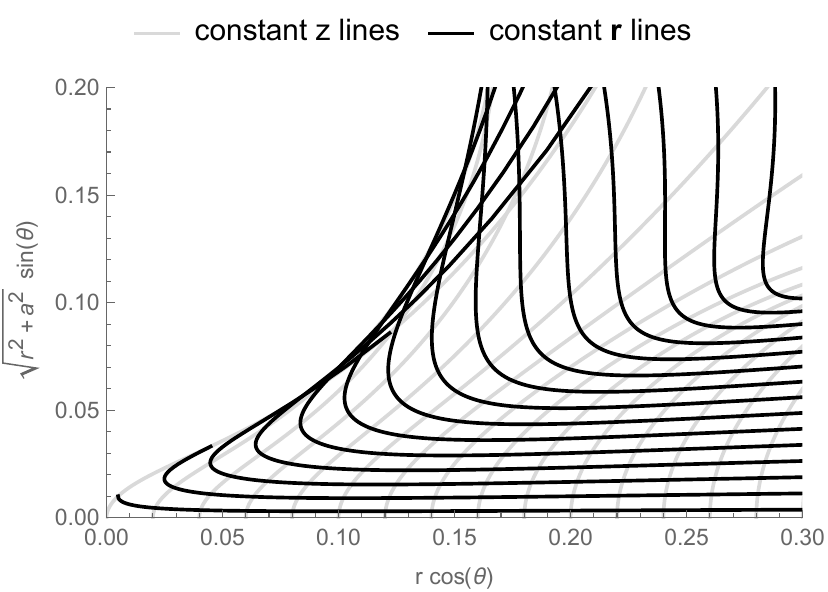}
  \caption{$\scriptstyle \text{Deformed cylindrical coordinates}$}
  \label{figgs:sub2}
\end{subfigure}
\caption{Plots for $M=1,~a=0.1$. Deformed cylindrical coordinates $(\bm{r},z)$ cover the whole relevant range of $r\cos(\theta)$ axis only for $\bm{r}\ll a^2/M$, and break down away from this region.}
\label{figgs:test}
\end{figure}
See Figure \ref{figgs:test} for a comparison of the oblate spheroidal coordinates and deformed cylindrical coordinates. Introducing the deformed cylindrical coordinates into the metric \eqref{kerr}, we obtain\footnote{The change of variables \eqref{crdtrafo} is set up specifically such that there is no $dz d\bm{r}$ term at order $\mathcal{O}(\bm{r}^1)$, the scale factor in front of all spatial coordinates is homogeneous at $\mathcal{O}(\bm{r}^0)$, and there is no non-trivial $d\phi^2$ term at order $\mathcal{O}(\bm{r}^4)$.}
\begin{align}
\label{gr2}
-d\tau^2&=-\frac{a^2+z (2 M+z)}{a^2+z^2}dt^2 +\frac{a^2+z^2}{a^2+z (2 M+z)}\left(d\bm{r}^2+\bm{r}^2d\phi^2+dz^2\right)\\
&+\bm{r}^2M\left(\frac{ \left(z^2 (M+z)-a^2 (M+3 z)\right)dt^2+4 a   z\left(a^2+z^2\right)dt d\phi}{\left(a^2+z^2\right)^2
   \left(a^2+z (2 M+z)\right)}+\frac{ z^3-a^2 (2 M+3 z)}{\left(a^2+z (2 M+z)\right)^3}dz^2 \right.\notag\\
	&\left.~~~~~~~~~~~~+\frac{ \left(a^4 (M+6 z)+2 a^2 z^2 (5 M+2 z)-z^4 (3 M+2
   z)\right)}{\left(a^2+z^2\right) \left(a^2+z (2 M+z)\right)^3}d\bm{r}^2\right)+dz d\bm{r}\,\mathcal{O}(\bm{r}^3)\notag\\
	&+\bm{r}^4\frac{2 a M (a-z) (a+z) \left(a^2 (M+3 z)+3 z^2
   (M+z)\right)}{\left(a^2+z^2\right)^3 \left(a^2+z (2 M+z)\right)^2}dt d\phi+\mathcal{O}(\bm{r}^4).\notag
\end{align}
The first line in \eqref{gr2} contains the leading order metric and features homogeneous scaling of generic cylindrical spatial coordinates by a $z$-dependent factor. The subsequent lines in \eqref{gr2} contain sub-leading metric terms sufficient to ensure that the overall metric remains a vacuum  solution of the Einstein field equations \eqref{EFE} (with $\Lambda=0$) in the vicinity ($\frac{\bm{r}}{a^2/M}\to 0$) of the trajectory of interest to order $\mathcal{O}(\bm{r}^2)$:
\begin{align}
\label{EFEr2}
R_{\mu\nu}-\frac{R}{2}g_{\mu\nu}=\mathcal{O}(\bm{r}^2).
\end{align}
Note in particular, that a $dz d\bm{r}\,\mathcal{O}(\bm{r}^3)$ term as well as several $\mathcal{O}(\bm{r}^4)$ terms are neglected in \eqref{gr2}, since they contribute to \eqref{EFEr2} at order $\mathcal{O}(\bm{r}^2)$ or higher and are therefore irrelevant for our purposes.

The specific constraint $\bm{r}\ll\frac{a^2}{M}$ arises due to the fact that the maximum (absolute) values of coefficients of $dt^2,~d\bm{r}^2$ and $dz^2$ terms at order $\mathcal{O}(\bm{r}^2)$ in \eqref{gr2} behave as $\propto\frac{M^2}{a^4}$ (found in the $z\to0$ region). Requiring $\bm{r}\ll\frac{a^2}{M}$ therefore suppresses these terms sufficiently via the $\bm{r}^2$ factor in front (making their scale $\ll\mathcal{O}(1)$), guaranteeing that the $\mathcal{O}(\bm{r}^0)$ metric terms are in fact of leading order on the entire range $0\leq z <\infty$.

What is the utility of formulating the metric as in \eqref{gr2}? 

By the first postulate of General Relativity, the relativity principle, a point-like observer at any point in any space-time perceives the metric in their immediate vicinity as the flat Minkowski metric, such that local physics is governed by Special Relativity. In other words, at any point in any physical space-time one can find normal coordinates, such that the leading term is given by the flat Minkowski metric, and first derivatives of the metric with respect to all coordinates vanish. Therefore in our case of interest, a point-like observer traveling from the vicinity of the singularity to infinity along a radial trajectory finds themselves in such normal coordinates at any given instant. If we desire to keep track of how the normal coordinates at a given point relate to the respective normal coordinates at a different point visited previously, a fruitful course of action is to relax the condition of normality for the single coordinate aligned with the path of trajectory, and by doing so treat coordinates at all visited points on an equal footing. The form of the metric \eqref{gr2} achieves precisely that, while maintaining homogeneous scaling of spatial coordinates at leading order in small $\bm{r}\ll a^2/M$. Normality is relaxed for the $z$-direction, which parametrizes different points on the trajectory, while normality is maintained in $t,\bm{r},\phi$ coordinates (perturbatively in $\bm{r}\ll a^2/M$, and up to constant rescaling at fixed $z$), with their first derivatives vanishing locally. This is analogous to the situation in the FLRW metric of cosmology, where the scale factor $\bm{a}(t)$ breaks the normality of the metric in the $t$-direction, while the rest of the coordinates are normal (up to constant rescaling at fixed $t$), and spatial coordinates experience homogeneous scaling.

So far, we have chosen a particular parametrization \eqref{gr2} for the space-time metric in the vicinity of a radial trajectory. To compare this metric with a typical cosmological metric of an unperturbed universe, we evaluate the metric along the radial trajectory parametrized in conformal time $\eta$. 

Due to the asymptotic flatness of the Kerr space-time, all sub-leading terms in \eqref{gr2} are guaranteed to vanish for increasing $z$ values. We therefore concentrate on the leading order metric in \eqref{gr2} to model the observable universe, and transform the time variable to conformal time $t = f(\eta)$, $dt = f'(\eta) d\eta$:
\begin{align}
\label{gfeta}
-d\tau^2=&-\frac{a^2+z (2 M+z)}{a^2+z^2}f'(\eta)^2d\eta^2 +\frac{a^2+z^2}{a^2+z (2 M+z)}\left(d\bm{r}^2+\bm{r}^2d\phi^2+dz^2\right)+\mathcal{O}(\bm{r}^2),
\end{align}
with the appropriate function $f(\eta)$ to be determined. In these variables, the zero'th component of the geodesic equation \eqref{geodesic} reads
\begin{align}
\label{geoEq0}
\frac{d^2 x^0}{d\tau^2}+\Gamma^0{}_{\alpha\beta}\frac{dx^\alpha}{d\tau}\frac{dx^\beta}{d\tau} = \frac{2 M (a^2-z^2) \eta '(\tau ) z'(\tau )}{\left(a^2+z^2\right) \left(a^2+z (2 M+z)\right)}+\frac{f''(\eta ) \eta '(\tau )^2}{f'(\eta )}+\eta
   ''(\tau ) = 0\,,
\end{align}
where we used that for metric \eqref{gfeta} the only non-vanishing components of $\Gamma^0{}_{\alpha\beta}$ are
\begin{align}
\Gamma^0{}_{00}=\frac{f''(\eta)}{f'(\eta)}~~~,~~~\Gamma^0{}_{03}=\Gamma^0{}_{30}=\frac{M(a^2-z^2)}{(a^2+z^2)\left(a^2+z(2M+z)\right)}\,,
\end{align}
and we seek the simplest solution $x^\mu=(\eta(\tau),0,0,z(\tau))$ that stays aligned with the $z$-axis.
Dividing \eqref{geoEq0} by $\eta'(\tau)$, we notice that
\begin{align}
\frac{\eta''(\tau)}{\eta'(\tau)}=\partial_\tau\log(\eta'(\tau))~~~,~~~&\frac{f''(\eta)\eta'(\tau)}{f'(\eta)}=\partial_\tau\log(f'(\eta))\,,\\
\frac{2 M (a^2-z^2)  z'(\tau )}{\left(a^2+z^2\right) \left(a^2+z (2 M+z)\right)}&=\partial_\tau\log\left(\frac{a^2+z(2M+z)}{a^2+z^2}\right)\,.
\end{align}
This means that after integrating in $\tau$ once, we have
\begin{align}
\log\left(\frac{a^2+z(2M+z)}{a^2+z^2}\right)+\log(f'(\eta))+\log(\eta'(\tau))=\log(\mathcal{C}),
\end{align}
where $\log(\mathcal{C})$ is an integration constant. Finally, we can solve for $\eta '(\tau )$ via subtraction and exponentiation, resulting in
\begin{align}
\label{vacetap}
\eta '(\tau )=\frac{\mathcal{C} \left(a^2+z(\tau )^2\right)}{\left(a^2+z(\tau ) (2 M+z(\tau ))\right) f'(\eta (\tau ))}\,.
\end{align}
To find $z'(\tau)$ in the case of a massive  particle, we use \eqref{gfeta} with $\bm{r}'(\tau)^2=0,\,\bm{r}^2\phi'(\tau)^2=\mathcal{O}(\bm{r}^2)$, plug in \eqref{vacetap}, and solve selecting the positive square root solution
\begin{align}
-1=&-\frac{a^2+z (2 M+z)}{a^2+z^2}f'(\eta)^2\eta'^2 +\frac{a^2+z^2}{a^2+z (2 M+z)}z'^2~~~\Rightarrow~~~z'(\tau)=\sqrt{\mathcal{C}^2-1-\frac{2 M z(\tau )}{a^2+z(\tau )^2}}.\notag
\end{align}
Note that in order for $z'(\tau)$ to be real valued on the whole range of $0\leq z(\tau)\leq\infty$, the integration constant $\mathcal{C}$ must satisfy $\mathcal{C}\geq \sqrt{1+M/a}$. Therefore, we can set $\mathcal{C}$ to e.g.\ the minimum admissible value 
\begin{align}
\label{vaczpt}
\mathcal{C}=\sqrt{1+\frac{M}{a}}~~~~~~\Rightarrow~~~~~~z'(\tau)=\sqrt{\frac{M}{a}-\frac{2 M z(\tau )}{a^2+z(\tau )^2}}.
\end{align}
Finally, to obtain $z'(\eta)$ we use the chain rule $z'(\tau)=z'(\eta)\eta'(\tau)$, plug in (\ref{vacetap}, \ref{vaczpt}) and solve
\begin{align}
\label{vaczpM}
z_m'(\eta)=\sqrt{\frac{M (a-z_m)^2}{(a+M) \left(a^2+z_m^2\right)}}\frac{a^2+z_m (2 M+z_m) }{a^2+z_m^2}f'(\eta ),
\end{align}
where we used the subscript $m$ to emphasize that this is the solution for a massive particle.

To find $z'(\eta)$ in case of a massless particle, we use \eqref{gfeta} with $d\bm{r}=0,\,d\phi=0,\,d\tau=0$, choose the positive square root solution and obtain
\begin{align}
\label{vaczp}
z_\gamma'(\eta)=\frac{a^2+z_\gamma (2 M+z_\gamma) }{a^2+z_\gamma^2}f'(\eta ),
\end{align}
where the subscript $\gamma$ emphasizes that this is the solution for a massless particle.

Recall that we introduced the change of time coordinate $t=f(\eta)$ in \eqref{gfeta} in order for $\eta$ to be conformal time. We can enforce this when we evaluate the metric on the trajectory parametrized by $\eta$, such that $z=z(\eta)$. In conformal time, we expect e.g.\ the equality $-g_{00}=g_{33}$ between metric components. Therefore, choosing the positive square root solution, we define $f(\eta)$ such that
\begin{align}
\label{solfp}
-g_{00}=\frac{a^2+z (2 M+z)}{a^2+z^2}f'(\eta)^2=\frac{a^2+z^2}{a^2+z (2 M+z)}=g_{33}~~~\Rightarrow~~~f'(\eta)=\frac{a^2+z(\eta)^2}{a^2+z(\eta) (2 M+z(\eta))}.
\end{align}
We will see shortly that to model an effective inflation alternative we should take the limit $a\ll M$. In that limit the massive and massless relations \eqref{vaczpM} and \eqref{vaczp} become asymptotically equivalent for large $z$ (or as soon as $z\sim\mathcal{O}(M)$), meaning that massive particles are accelerated close to the speed of light after they escape towards infinity. Therefore, in the following we focus on the massless case without loss of generality. 

Plugging \eqref{solfp} into \eqref{vaczp}, we observe
\begin{align}
\label{solzp}
z'(\eta)= 1\,,~~~\text{the simplest solution being}~~~z(\eta)=\eta.
\end{align}
The coordinate $\phi$ is compact, and coordinate $\bm{r}$ is already local to the trajectory for all $\eta$. With the solution \eqref{solzp} in hand we can replace $z$ by a new axial coordinate that is similarly local to the trajectory performing the following coordinate transformation:
\begin{align}
\label{locz}
z = \eta + Z~~~,~~~dz = d\eta + dZ\,,
\end{align}
where the $\eta$ summand accounts for the global evolution in conformal time along the geodesic, while $Z$ is a new local axial coordinate. Being local means that $Z$ is of the same scale as $\bm{r}$:
\begin{align}
Z\sim \mathcal{O}(\bm{r})~~~\text{such that}~~~Z\ll a^2/M\,.
\end{align}
Thus, transformed to conformal time (with $f'(\eta)=\frac{a^2+\eta^2}{a^2+\eta (2 M+\eta)}$), evaluated on the trajectory of interest via \eqref{locz}, and expanded to leading order in small $Z\sim \mathcal{O}(\bm{r})$  (therefore, restricting to be local to the geodesic trajectory), the leading order metric \eqref{gfeta} reads
\begin{align}
\label{gloceta}
-d\tau_{\bm{r},\phi,Z\text{ local}}^2=&\frac{a^2+\eta^2}{a^2+\eta (2 M+\eta)}\left(2d\eta dZ+d\bm{r}^2+\bm{r}^2d\phi^2+dZ^2\right)+\mathcal{O}(\bm{r}^1),
\end{align}
where the off-diagonal elements appear due to our frame of reference and do not change the fact that the metric is conformally flat.

It may appear somewhat unsatisfactory that metric \eqref{gloceta} has off-diagonal terms. The main obstacle in diagonalizing \eqref{gloceta} is that spatial coordinates are local $\bm{r},\,Z\ll a^2/M$, or compact in case of $\phi$, while the conformal time coordinate $\eta$ is global. In the vicinity of any arbitrary but fixed conformal time $\eta=\eta_*> a^2/M$, we can consider local time evolution:
\begin{align}
\label{trafETAdiag}
\eta=\eta_*+\bm{\eta}~~~,~~~d\eta=d\bm{\eta}~~~\text{with local $\bm{\eta}\sim\mathcal{O}(\bm{r})$, such that}~~~\bm{\eta}\ll a^2/M.
\end{align}
This allows us to transform the axial coordinate $Z$ once more, without losing the local property:
\begin{align}
\label{trafZdiag}
Z=\bm{z}-\bm{\eta}~~~,~~~dZ=d\bm{z}-d\bm{\eta}~~~\text{with local $\bm{z}\sim\mathcal{O}(\bm{r})$, such that}~~~\bm{z}\ll a^2/M.
\end{align}
Implementing \eqref{trafETAdiag} and \eqref{trafZdiag} in \eqref{gloceta}, and expanding to leading order in local variables, we obtain the now completely local leading order metric:
\begin{align}
\label{glocall}
-d\tau_{\text{local}}^2=&\frac{a^2+\eta_*^2}{a^2+\eta_* (2 M+\eta_*)}\left(-d\bm{\eta}^2 +d\bm{r}^2+\bm{r}^2d\phi^2+d\bm{z}^2\right)+\mathcal{O}(\bm{r}^1).
\end{align}
Metric \eqref{glocall} is how a local observer perceives their conformally flat surroundings after having traveled along the axial geodesic for a conformal time $\eta_*> a^2/M$. Note especially the locally emerging spherical symmetry in the spatial coordinates, that was not yet present in \eqref{gr2} even at leading order, due to the homogeneous scaling factor of the spatial coordinates depending on the global $z$ coordinate.

\begin{figure}[tbp]
\centering 
\includegraphics[width=0.8\textwidth]{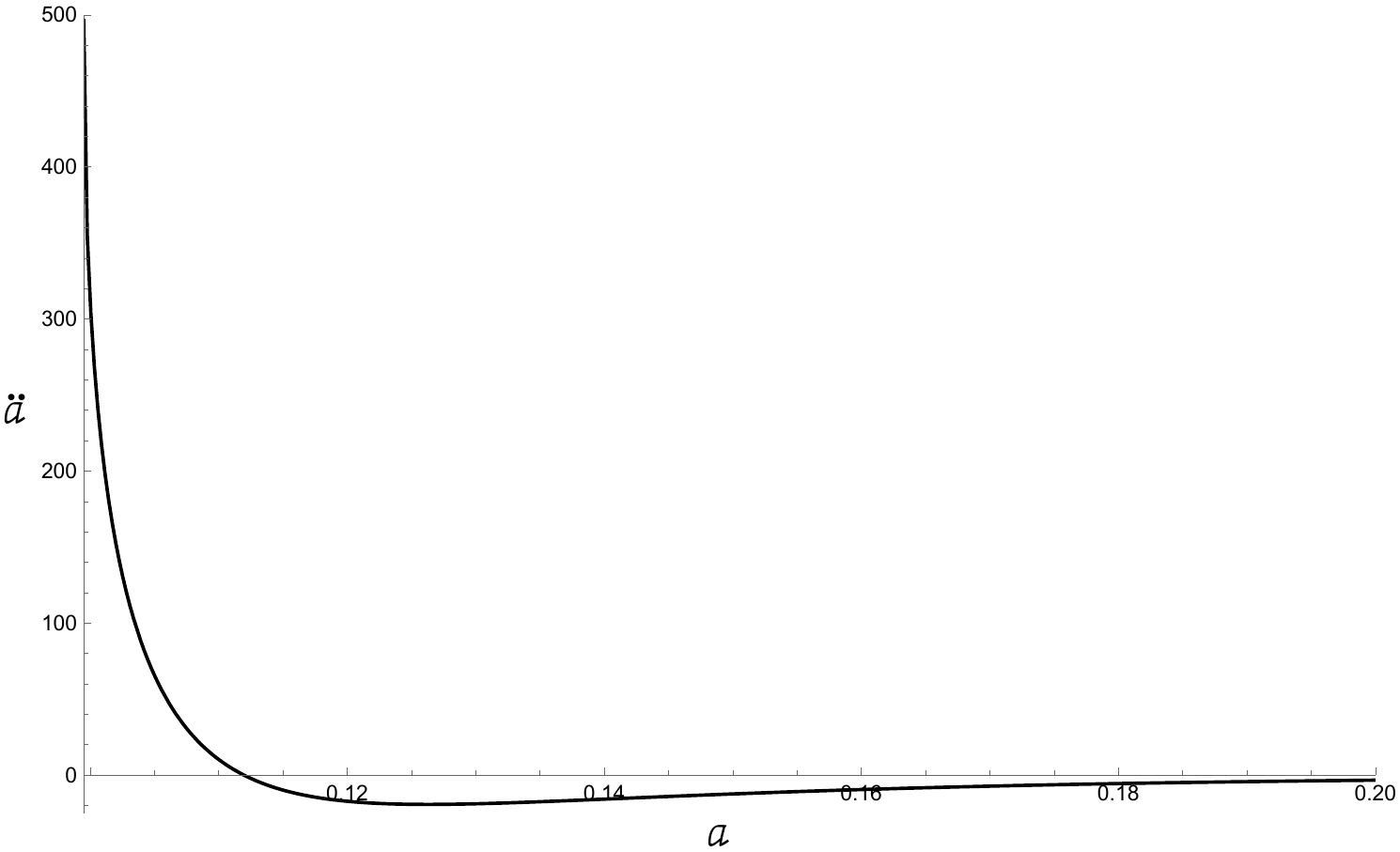}
\caption{ Acceleration $\frac{d^2 }{dt^2}\bm{a}$ of the scale factor \eqref{alead} as a function of the scale factor $\bm{a}$ itself, evaluated at $M=1,~a=10^{-1}$. The curve starts at $\bm{a}(a)=\sqrt{a/(a+M)}$, where it assumes its maximum value $\frac{1}{2}M a^{-3}\sqrt{a/(a+M)}$. After a brief period of positive acceleration, the expansion starts decelerating, until it asymptotically comes to a halt when the scale factor reaches unity  $\bm{a}(\infty)=1$.}
\label{comovingH}
\end{figure}

Metric \eqref{gloceta} is analogous to a conventional FLRW metric in conformal time (trivially transformed), with an effective conformal scale factor 
\begin{align}
\label{alead}
\bm{a}(\eta)=\sqrt{\frac{a^2+\eta^2}{a^2+\eta (2 M+\eta)}}.
\end{align}
The scale factor equals $\bm{a}(0)=1$ at the initial $z=0$ surface, and falls to a minimum of $\bm{a}(a)=\sqrt{{a}/({a+M})}$ when conformal time reaches the value of the Kerr spin parameter $\eta=a$. Subsequently, the scale factor starts growing and asymptotes back to $\bm{a}(\infty)=1$. There is a brief period of accelerated expansion at the beginning of the growth phase, see Figure \ref{comovingH} for a qualitative plot.\footnote{Note that particles on the geodesic at late time still keep accelerating outward according to the gravitational push of the negative mass singularity. The decrease in the acceleration of the scale factor on the other hand intuitively parametrizes the decrease of the gravitational field the farther we get from the singularity.} However, the fact that the scale factor starts at unity and decreases in magnitude initially is sufficient to ensure prior causal contact between regions reentering the horizon during the later expansion phase. The smaller the spin parameter $a$, the smaller the scale factor becomes at its lowest point. Therefore, we concentrate on the small spin case $0<a\ll M$ in the following, which ensures that any amount of relative expansion starting from the lowest point $\bm{a}(a)=\sqrt{{a}/({a+M})}$ can be accommodated. This demonstrates how non-trivial time evolution of the cosmological scale factor solving the horizon problem can be achieved dynamically e.g.\ by traversing the gravitational field created by a super-galactic ring singularity in the far past,\footnote{Super-galactic ring singularity here means that the entire observable universe at some time in the past was compressed to a region smaller than the radius of the ring singularity.} instead of the conventional approach of producing inflation by introducing space-filling energy-momentum at the particular time when the scale factor is evolving. 

Note that the contraction and expansion of the scale factor are purely local effects perceived by the observer due to passage of conformal time and their progress along the axial trajectory through the gravitational field of the singularity. It is consistent with observation insofar as it solves the horizon problem. However, the underlying space-time the observer is traversing is static at all times, undergoing no contracting nor expanding phases.

So far, no gravitating energy momentum content is traveling along the geodesic trajectory, so that the above metric evolves into the flat Minkowski metric for $\eta\to\infty$. In the next section we add gravitating radiation emerging from the initial $z=0$ surface, to model a radiation dominated unperturbed universe in this setting.

\section{Radiation Dominated Universe}
\label{sec:radiation}
In this section we introduce a gravitating radiation energy density emerging from the $z=0$ surface and flowing along the outgoing axial geodesic. An explanation for the origin of this radiation is as follows: We think of the radiation as arriving from infinity and axially falling into the rotating black hole in the positive mass region of the geometry, propagating through the \textit{`eye of the needle`} of the ring singularity, and into the negative mass black hole region of interest here. This trajectory naturally passes through the point $(z=0,\,\bm{r}=0)$, which is located on the axis of rotation of the ring singularity. 

We assume the radiation energy momentum flow to be constant at the time scale of interest, and are therefore looking for a static perturbative correction to the metric tensor. The ring singularity collapses to a point if the Kerr spin parameter $a$ is set to zero exactly, closing the passage through the \textit{'eye of the needle'} of the ring singularity, in which case the flow of radiation should vanish. Therefore we assume that the emerging radiation density is proportional to $a$:
\begin{align}
\label{rhozero}
\rho_{\gamma,0}\propto\frac{a}{M}.
\end{align}
Furthermore, as observed in the previous section, the smaller the Kerr spin parameter $a$ is, the more the effective scale factor \eqref{alead} contracts before the expansion phase. We can accommodate any relative amount of expansion from the lowest point if we consider the small spin parameter regime $0<a\ll M$.

Introducing a non-zero energy momentum flow of radiation causes the geometry of space-time to shift. In order to account for these changes, we take the following ansatz for the perturbation of metric \eqref{gr2}:
\begin{align}
\label{gr2def}
-d\tau^2&=-\frac{a^2+z (2 M+z)}{a^2+z^2}dt^2\left(1-\frac{a}{M}(1+\beta)\, h_1(z)+\frac{a \bm{r}^2}{M^3}h_3(z)\right) +2 dt dz \frac{a \bm{r}^2}{M^3} h_2(z)\\
&+\frac{a^2+z^2}{a^2+z (2 M+z)}\left(d\bm{r}^2+\bm{r}^2d\phi^2+dz^2\right)\left(1+\frac{a}{M} h_1(z)+\frac{a \bm{r}^2}{M^3}h_4(z)\right)\notag\\
&+\bm{r}^2M\left(\frac{ \left(z^2 (M+z)-a^2 (M+3 z)\right)dt^2+4 a   z\left(a^2+z^2\right)dt d\phi}{\left(a^2+z^2\right)^2
   \left(a^2+z (2 M+z)\right)}+\frac{ z^3-a^2 (2 M+3 z)}{\left(a^2+z (2 M+z)\right)^3}dz^2 \right.\notag\\
	&\left.~~~~~~~~~~~~+\frac{ \left(a^4 (M+6 z)+2 a^2 z^2 (5 M+2 z)-z^4 (3 M+2
   z)\right)}{\left(a^2+z^2\right) \left(a^2+z (2 M+z)\right)^3}d\bm{r}^2\right)+dz d\bm{r}\,\mathcal{O}(\bm{r}^3)\notag\\
	&+\bm{r}^4\frac{2 a M (a-z) (a+z) \left(a^2 (M+3 z)+3 z^2
   (M+z)\right)}{\left(a^2+z^2\right)^3 \left(a^2+z (2 M+z)\right)^2}dt d\phi+\mathcal{O}(\bm{r}^4),\notag
\end{align}
where $\beta$ is a constant and $h_1(z),h_2(z),h_3(z),h_4(z)$ are perturbation functions to be determined. We focus on the case where purely spatial deformations are homogeneous. As discussed above, each perturbation is accompanied by a factor of $\frac{a}{M}$, making sure that for $a=0$ the perturbations vanish along with the emerging radiation. We are seeking $h_i(z)$ with $i=1,2,3,4$ solutions of the Einstein field equations at leading order in $a/M\to0$ as well as $\frac{\bm{r}}{a^2/M}\to0$.  We assume that at leading order $\mathcal{O}(a^1,\bm{r^0})$ the metric deformation can be parametrized by a single function $h_1(z)$ similar to a scale factor; however, for more generality we allow a relative constant factor $-(1+\beta)$ in the $dt^2$ metric component. Note that we have introduced a $dt\,dz$ perturbation into the metric at the sub-leading order $\mathcal{O}(\bm{r}^2)$ to account for non-zero energy momentum flow in the $z$-direction without breaking the diagonal form of the leading order metric. 

Radiation \textit{fluid} typically possesses non-zero pressure. However, in our case we are instead interested in radiation propagating at the speed of light in the $z$-direction. This is equivalent to energy momentum of a superposition of plane waves and takes the familiar form
\begin{align}
\label{Tdef}
T^{\mu\nu}=\rho(z)U^\mu U^\nu\, ,\text{ with four-velocity }U^\mu=\left(\frac{1}{\sqrt{-g_{00}}},0,0,\frac{1}{\sqrt{g_{33}}}\right),
\end{align}
where we allow the energy density $\rho(z)$ to vary statically with $z$ along the geodesic trajectory. Here, the four-velocity $U^\mu$ is lightlike $g_{\mu\nu}U^\mu U^\nu=0$ for $\frac{\bm{r}}{a^2/M}\to 0$ by construction. 

At leading order in $\frac{\bm{r}}{a^2/M}\to0$, the covariant derivative vanishing condition \eqref{covT} leads to a single non-trivial equation:
\begin{align}
\nabla_\nu T^{0\nu}\propto \nabla_\nu T^{3\nu}\propto  \partial_z \left[\rho (z) \left(a h_1(z)+M\right) \left(M-a (1+\beta) h_1(z)\right)\right]= 0.
\end{align}
This directly integrates to\footnote{Note that $\rho$ has units of $[\rho]=L^{-2}$ as appropriate in the geometrized units convention.}
\begin{align}
\label{rhorad}
\rho=\frac{\rho _{\gamma,0 }}{\left(a h_1+M\right) \left(M-a (1+\beta) h_1\right)}=\frac{a}{M\left(a h_1+M\right) \left(M-a (1+\beta) h_1\right)},
\end{align}
where the initial energy density \eqref{rhozero} appears as an integration constant. 
Making use of \eqref{rhorad}, the other two components of \eqref{covT} vanish or are sub-leading in $a\ll M$
\begin{align}
\nabla_\nu T^{1\nu}=0\,,~~~\nabla_\nu T^{2\nu}=\mathcal{O}(a^2,\bm{r}^0).
\end{align}
We utilize \eqref{gr2def}, \eqref{Tdef} and \eqref{rhorad} in the Einstein field equations \eqref{EFE}. The single non-trivial off-diagonal equation expanded to leading order $\mathcal{O}(a^1,\bm{r}^0)$ fixes $h_{2}$ in terms of other quantities
\begin{align}
\label{solh2}
R^{03}-\frac{R}{2}g^{03} = T^{03}~~~~~\Rightarrow~~~~~h_{2}=\frac{ z }{  4 M+2 z}.
\end{align}
All other off-diagonal elements either vanish exactly, or are of sub-leading order $\mathcal{O}(a^2,\bm{r}^0)$ and are therefore irrelevant for our purposes.
 
The diagonal elements $R^{\mu\mu}-\frac{R}{2}g^{\mu\mu} = T^{\mu\mu}$ (no summation over $\mu$) of the Einstein equations \eqref{EFE} at order $\mathcal{O}(a^1,\bm{r}^0)$ (and an order $\mathcal{O}(a^1,\bm{r}^{-2})$ term for $\mu=3$) then lead to the following three non-trivial linearly independent constraints
\begin{align}
\label{req0}
0=&M^3  (3 M+4 z)h_1-z (2 M+z) \left(M^3 h_1'+z^2+ (2 M+z)\left(M^2 z
   h_1''+4 z  h_4\right)\right),\\
\label{req1}
0=&z (2 M+z) \left((2 M+z)\left[2 z \left(h_3+h_4\right)-\beta M^2 z  h_1''\right] +2 (1+\beta) M^3 h_1'\right)\notag\\
	&-2 M^3 h_1 ((1+\beta) (M+z)+z),\\
\label{req2}
0=&z (2 M+z) \left(z \left(2 \left(h_3+h_4\right) (2
   M+z)-z\right)-(1+\beta) M^3 h_1'\right)\notag\\
	&-M^3 h_1 ((2 \beta+5) M+2 (2+\beta) z).
\end{align}
Note that no derivatives of $h_3$ and $h_4$ functions make an appearance, which allows us to simply solve two out of the three equations algebraically:
\begin{align}
\label{solh3h4}
h_3=&\frac{z (2 M+z) \left(z \left((2 \beta+1)
   M^2 (2 M+z) h_1''+z\right)-(4 \beta+3) M^3 h_1'\right)+M^3 h_1 ((4 \beta+1) M+4 (1+\beta) z)}{4 z^2 (2
   M+z)^2}\,,\notag\\
h_4=&\frac{M^3 h_1 (3 M+4 z)-z (2 M+z) \left(M^3 h_1'+z \left(M^2 (2
   M+z) h_1''+z\right)\right)}{4 z^2 (2 M+z)^2}.
\end{align}
The third equation then reduces to a differential equation in $h_1$ only:
\begin{align}
\label{diffeqh1}
0=z (2 M+z) \left(z \left(z-\beta M^2 (2 M+z) h_1''\right)+3 (1+\beta) M^3
   h_1'\right)+3 M^4 h_1.
\end{align}
Despite its simple appearance, this equation is rather complicated and, according to \textit{Mathematica}, its general solution can be given in terms of integrals over hypergeometric functions
\begin{align}
h_1=&z^{\frac{5 \beta -R_1+3}{4 \beta }} (2 M+z)^{\frac{R_2-\beta -3}{4
   \beta }}\left[z^{\frac{R_1}{2 \beta }} \, _2F_1\left({\scriptstyle{\frac{R_1+R_2}{4
   \beta },\frac{R_1+R_2}{4 \beta }+1}\atop{\frac{R_1}{2 \beta
   }+1}};-\frac{z}{2 M}\right)\times\right.\\
\times& \left(c_1+\int _a^z dx\frac{8(7
   \beta  (\beta +2)+3) x^{\frac{3 \beta
   -R_1-3}{4 \beta }} (2 M+x)^{\frac{3-3 \beta -R_2}{4 \beta }}  }{M H(x)}\, _2F_1\left({\scriptstyle{\frac{R_2-R_1}{4 \beta },\frac{ R_2-R_1}{4 \beta }+1}\atop{1-\frac{R_1}{2 \beta }}};-\frac{x}{2
   M}\right)\right)\notag\\
	+&\, _2F_1\left({\scriptstyle{\frac{R_2-R_1}{4 \beta
   },\frac{R_2-R_1}{4 \beta }+1}\atop{1-\frac{R_1}{2 \beta }}};-\frac{z}{2
   M}\right)\times\notag\\
\times&\left.\left(c_2-\int _a^z dx\frac{8 (7 \beta  (\beta +2)+3) x^{\frac{3 \beta +R_1-3}{4 \beta }}
   (2 M+x)^{\frac{3-3 \beta -R_2}{4 \beta }}  }{M
   H(x)}\,
   _2F_1\left({\scriptstyle{\frac{R_1+R_2}{4 \beta },\frac{R_1+R_2}{4 \beta
   }+1}\atop{\frac{R_1}{2 \beta }+1}};-\frac{x}{2 M}\right)\right)\right],\notag
\end{align}
where integration constants $c_1,\,c_2$ can be used to alter the lower boundaries of integration on each of the two integrals respectively.\footnote{By default we have chosen lower boundaries of integration to be equal to the Kerr spin parameter $a$ for $c_1=c_2=0$, such that the metric deformation is smallest at the minimum of the effective scale factor \eqref{alead}.} Abbreviations $R_1,\,R_2$ denote
\begin{align}
R_1=\sqrt{\beta  (25 \beta +42)+9}~~~,~~~R_2=\sqrt{\beta  (\beta +18)+9}\,,
\end{align}
 and the function $H(x)$ is given by
\begin{align}
H&(x)=x \Big(\beta  \left(\beta  \left(8 \beta -3 R_1+7 R_2+8\right)+2
   \left(7 R_2-5 R_1\right)\right)+3 \left(R_2-R_1\right)\Big) \times\\
	\times&\,
   _2F_1\left({\scriptstyle{\frac{R_2-R_1}{4 \beta }+1,\frac{R_2-R_1}{4
   \beta }+2}\atop{2-\frac{R_1}{2 \beta }}};-\frac{x}{2 M}\right) \,
   _2F_1\left({\scriptstyle{\frac{R_1+R_2}{4 \beta },\frac{R_1+R_2}{4 \beta
   }+1}\atop{\frac{R_1}{2 \beta }+1}};-\frac{x}{2 M}\right)\notag\\
	+&\,
   _2F_1\left({\scriptstyle{\frac{R_2-R_1}{4 \beta },\frac{4 \beta -R_1+R_2}{4 \beta
   }}\atop{1-\frac{R_1}{2 \beta }}};-\frac{x}{2 M}\right) \left[4 (7 \beta  (\beta
   +2)+3) R_1 M  \, _2F_1\left({\scriptstyle{\frac{R_1+R_2}{4 \beta },\frac{R_1+R_2}{4 \beta }+1}\atop{\frac{R_1}{2 \beta }+1}};-\frac{x}{2 M}\right)\right.\notag\\
	-&\left.x \Big(\beta 
   \left(\beta  \left(8 \beta +3 R_1+7 R_2+8\right)+2 \left(5 R_1+7
   R_2\right)\right)+3 \left(R_1+R_2\right)\Big) \, _2F_1\left({\scriptstyle{\frac{R_1+R_2}{4 \beta }+1,\frac{R_1+R_2}{4 \beta
   }+2}\atop{\frac{R_1}{2 \beta }+2}};-\frac{x}{2 M}\right)\right].\notag
\end{align}
In order for the perturbations to make physical sense, the sub-leading terms $h_2,\,h_3,\,h_4$ may not grow large for $z\to\infty$, while the leading term $h_1$ is expected to grow with $z$. The degenerate case $\beta=0$ causes the $h_1$ solution to scale with a power of $z$ not consistent with the expectation of a radiation dominated universe. Therefore, we discard that case as unphysical. Making use of standard hypergeometric expansions as well as the initial differential equation \eqref{diffeqh1}, it is straightforward to determine that for $\beta\neq 0$:
\begin{align}
\label{h1zinf}
h_1|_{z\to\infty}=\frac{1}{2M^2\beta}z^2+{{\scriptstyle\mathcal{O}}}(z^2)\,,
\end{align}
while equations \eqref{solh2} and \eqref{solh3h4} then reveal
\begin{align}
h_2|_{z\to\infty}&=\frac{1}{2}+{{\scriptstyle\mathcal{O}}}(z^0)\,,\notag\\
\label{h234zinf}
h_3|_{z\to\infty}&=\frac{1}{4}\left(3+\frac{1}{\beta}\right)+{{\scriptstyle\mathcal{O}}}(z^0)\,,\\
h_4|_{z\to\infty}&=-\frac{1+\beta}{4\beta}+{{\scriptstyle\mathcal{O}}}(z^0)\,,\notag
\end{align}
where the small ${{\scriptstyle\mathcal{O}}}(x^n)$ notation denotes terms that grow strictly slower than $x^n$. For the leading order perturbation to be positive at large $z$ \eqref{h1zinf}, we must enforce $\beta>0$. The sub-leading perturbations \eqref{h234zinf} indeed do not grow with $z$, as required.

Note that it would not be meaningful to investigate the behavior of $h_1,\,h_2,\,h_3,\,h_4$ at the opposite end of the range $z\to 0$, since all above results are obtained perturbatively at order $\mathcal{O}(a)$ in small $\frac{a}{M}$ expansion. The limit $z\to 0$ would probe $\frac{z}{M}<\frac{a}{M}$ no matter how small $\frac{a}{M}$ is, while our results are valid only for $\frac{z}{M}\gg \frac{a}{M}$ due to their perturbative nature. Nevertheless, we find that even at a $z$ value as low as $z=a$ the perturbation functions are appropriately well behaved:
\begin{align}
h_1|_{z=a}&=0\,,~~~~~~~~~~~~~~~~~~~~~~~h_2|_{z=a}=\frac{a}{4M+2a}\,,\\
h_3|_{z=a}&=\frac{a(1+3\beta)}{4(2M+a)\beta}\,,~~~~~~~~h_4|_{z=a}=-\frac{a(1+\beta)}{4(2M+a)\beta}\,.
\end{align}

Next, we evaluate the metric on the radial geodesic trajectory, transformed to locally conformal time $t=f(\eta),\,dt=f'(\eta)d\eta$. As in the previous section, we focus on the leading order metric only to model the observable universe
\begin{align}
\label{gfetah1}
-d\tau^2=&-\frac{a^2+z (2 M+z)}{a^2+z^2}f'(\eta)^2\left(1-\frac{a}{M}(1+\beta)\, h_1(z)\right)d\eta^2 \\
&+\frac{a^2+z^2}{a^2+z (2 M+z)}\left(1+\frac{a}{M} h_1(z)\right)\left(d\bm{r}^2+\bm{r}^2d\phi^2+dz^2\right)+\mathcal{O}(\bm{r}^2).\notag
\end{align}
To find $z'(\eta)$ in case of a massless particle, we use \eqref{gfetah1} with $d\bm{r}=0,\,d\phi=0,\,d\tau=0$, choose the positive square root solution and obtain
\begin{align}
\label{zph1}
z'(\eta)=\frac{a^2+z (2 M+z) }{a^2+z^2}f'(\eta )\sqrt{\frac{1-\frac{a}{M}(1+\beta)\, h_1(z)}{1+\frac{a}{M} h_1(z)}}.
\end{align}
The relative minus sign in the square root looks like it may pose a problem. It can be traced back all the way to the $(1-\frac{a}{M}(1+\beta)\, h_1(z))$ factor in the $dt^2$ component of \eqref{gr2def}. We can think of that factor as the first two terms in a small $a/M$ expansion of e.g.\ $1/(1+\frac{a}{M}(1+\beta)\, h_1(z))$, which remedies the relative minus sign.

Solving $-g_{00}=g_{33}$ for $f'(\eta)$ when we evaluate the metric on the trajectory $z=z(\eta)$ as in \eqref{solfp}, we find
\begin{align}
\label{solfph1}
f'(\eta )=\frac{a^2+z^2}{a^2+z (2 M+z)}\sqrt{\frac{1+\frac{a}{M} h_1(z)}{1-\frac{a}{M}(1+\beta)\, h_1(z)}}.
\end{align}
Plugging \eqref{solfph1} into \eqref{zph1}, the relation simplifies analogously to \eqref{solzp}:
\begin{align}
z'(\eta)= 1\,,~~~\text{the simplest solution being}~~~z(\eta)=\eta.
\end{align}
Once again, we can introduce a new axial coordinate $Z$ local to the trajectory (so that $Z\sim \mathcal{O}(\bm{r})$) via \eqref{locz}.
Therefore, transformed to conformal time and evaluated on the trajectory of interest, the perturbed local leading order metric reads
\begin{align}
\label{gloctauh}
-d\tau^2_{\bm{r},\phi,Z\text{ local}}=&\frac{a^2+\eta^2}{a^2+\eta (2 M+\eta)}\left(1+\frac{a}{M} h_1(\eta)\right)\left(2d\eta dZ+d\bm{r}^2+\bm{r}^2d\phi^2+dZ^2\right)+\mathcal{O}(\bm{r}^1).
\end{align}
Analogously to the previous section, transformations \eqref{trafETAdiag} and \eqref{trafZdiag} can be used to diagonalize the metric locally, if desired.

\begin{figure}[tbp]
\centering 
\includegraphics[width=0.9\textwidth]{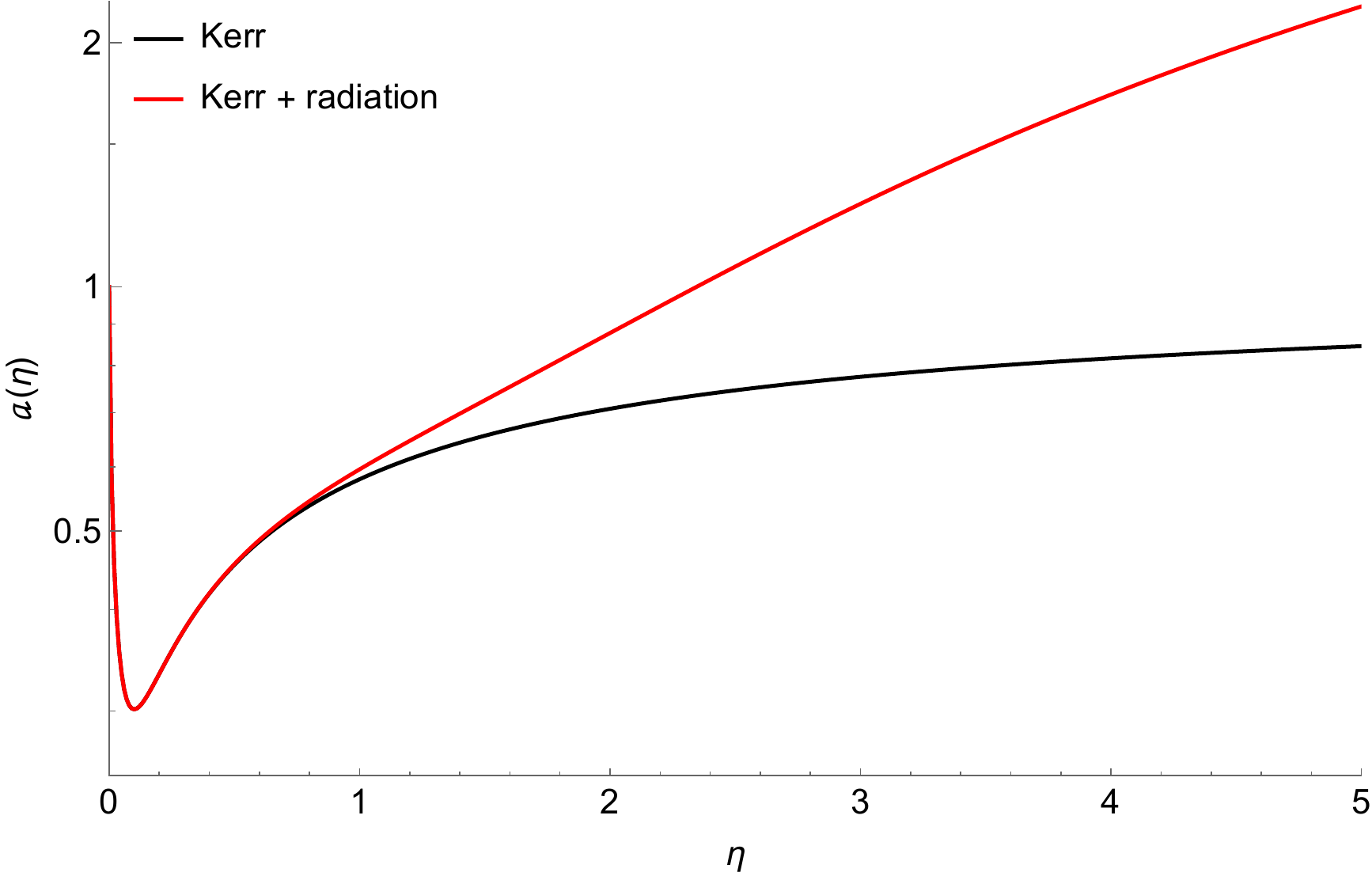}
\caption{ Comparison of effective scale factors \eqref{alead} and \eqref{aleadh1}, for demonstration evaluated at $c_1=c_2=0,\,M=1,\,a=10^{-1},\,\beta=1$. Note that, due to only a single order of magnitude difference in $M$ and $a$ for visualization purposes, the radiation contribution appears dramatically exaggerated.}
\label{aVSah1}
\end{figure}

Due to the introduction of the gravitating flow of radiation along the radial geodesic, the effective scale factor \eqref{alead} is deformed to
\begin{align}
\label{aleadh1}
\bm{a}(\eta)=&\sqrt{\frac{a^2+\eta^2}{a^2+\eta (2 M+\eta)}\left(1+\frac{a}{M} h_1(\eta)\right)}
\end{align}
in \eqref{gloctauh}. From \eqref{h1zinf} we see that for late $\eta$ this implies
\begin{align}
\label{aasymp}
\bm{a}(\eta)|_{\eta\gg M}\approx \left(\frac{a}{2\beta M^3}\right)^{\frac{1}{2}}\eta.
\end{align}
For a radiation dominated universe we expect $\bm{a}(t)\propto t^{1/2}$ and the transformation to conformal time therefore requires $t'(\eta)^2\propto t(\eta)$, which is solved by $t(\eta)\propto \eta^2$. The result \eqref{aasymp} is thus precisely consistent with our expectation for a radiation dominated universe. The free parameter $\beta$ can be used to fit the proportionality factor to observation. See Figure \ref{aVSah1} for a visualization of the effective scale factor.

\section{Conclusion and Outlook}
\label{sec:conclusion}
In this work we have explored the scenario that our observable universe has been emitted from the vicinity of a negative mass ring singularity in a Kerr space-time, and is traveling along a radial geodesic parallel to the axis of rotation of the ring singularity. In appropriate cylindrical coordinates $\bm{r},z$, assuming large enough spin and mass parameters $a$ and $M$ such that the observable universe is well described by the leading terms in an expansion around small $\frac{\bm{r}}{a^2/M}\to 0$, we have shown that the metric on the trajectory exhibits a particular conformal scale factor. The inherent contraction and expansion properties of the scale factor solve the horizon problem. Subsequently, we introduced a gravitating stream of radiation flowing statically along the trajectory, to model a radiation dominated universe. With an appropriate ansatz for metric perturbation, we then demonstrated that the resulting asymptotic growth of the scale factor indeed matches the expected result in a radiation dominated universe. As the features stemming from the singularity in the far past die off asymptotically at large conformal time, the transition to a matter dominated universe could then be made utilizing the conventional FLRW metric.

One objection to working with the negative mass Kerr space-time is the existence of unbounded modes in linear time-dependent metric perturbations \cite{Dotti:2022djh,Cardoso:2006bv}, which make the space-time unstable. However, just the presence of an instability alone does not preclude a meta-stable state that may have a life-time of the age of the universe. Additionally, the fact that we are specifically working with only a patch of the space-time in the vicinity of the trajectory along the axis of rotation means that, especially in the case when radiation is present, a different global completion of the metric may exist that does not exhibit the same instability. Furthermore, metric perturbations as introduced in \eqref{gr2def} are static, while \cite{Dotti:2022djh,Cardoso:2006bv} find instability only due to time-dependent perturbations. Therefore, perturbations \eqref{gr2def} themselves do not pose any problems in this regard.

The sub-leading terms in metric \eqref{gr2} vanish (or asymptote to a constant in metric \eqref{gr2def}, negligible with respect to the growing leading terms) for large $z$, which is achieved for late conformal time $\eta$ on the trajectory. However, at early conformal time $\eta\approx 0$ the sub-leading terms can be relevant away from the absolute center of the trajectory. Therefore, it would be interesting to investigate if the sub-leading terms can imprint some signature onto the streaming radiation at early times, which could be measured at late times. Technically, the initial change of variables \eqref{crdtrafo} is not completely unique. Choosing a slightly different coordinate transformation can change the early conformal time behavior of sub-leading terms and therefore potentially imprint different signatures onto the radiation. Similarly, a re-parametrization of the $z$-coordinate to any monotonically growing function starting at zero and linearly asymptoting to infinity may possibly change the quantitative details but not the qualitative picture.

Furthermore, the ansatz of metric perturbation \eqref{gr2def} is physically motivated with the target form of the resulting leading order metric in mind, but not unique. It would be interesting to explore modifications of the ansatz, to potentially determine an exhaustive equivalence class of qualitatively similar perturbative solutions.

Regarding over-densities observed in the Cosmic Microwave Background (CMB), there does not seem to be any reason to rule out sub-leading density fluctuations in a generic stream of radiation propagating through spacetime. In fact, without a singular, extremely hot and dense point of origin such as The Big Bang enforcing extreme homogeneity and symmetry conditions, an assumption of perfect homogeneity of a stream of radiation would be hard to justify. During the contraction phase of the effective conformal scale factor, the energy density of the radiation rises, shrinking the scale of inhomogeneity and making the universe appear more homogeneous and isotropic at that conformal time. The subsequent expansion of the effective conformal scale factor then reverses this effect.

Even though the Kerr spacetime has cylindrical symmetry, each local observer finds themselves in normal coordinates that by definition are spherically symmetric. With \eqref{glocall} and \eqref{gloctauh} we found that for a local observer traveling along the axial geodesic these normal coordinates relate to each other by a change of the conformal scale factor in conformal time $\eta$, and spatial deviations from spherical symmetry occur at scales $\mathcal{O}(\frac{a^2}{M})$. Parameters $a$ and $M$ can be chosen such that scale $\frac{a^2}{M}$ is much larger than the observable universe. Therefore, local evolution of perturbations into overdensities and galaxy clusters can appear almost perfectly spherically symmetric to a local observer. Nevertheless, depending on the particular realization of $a$ and $M$ values, it may be possible that a sub-leading cylindrical breaking of spherical symmetry is present. Should even the tiniest indication of a preferred direction ever be observed on cosmological scale, it would provide experimental support to the scenario outlined in this work.

Furthermore, in this scenario any potential sub-leading density perturbations within the radiation are not down-stream consequences of fluctuations in any other field, but rather are initial conditions that we are free to set at will. Therefore, this scenario does not provide any constraints in this regard. However, due to the fact that no violent initial event such as The Big Bang is required in this scenario, we do not expect any primordial gravitational waves to be present. Should the detection of primordial gravitational waves ever be confirmed successfully, it will rule out the scenario discussed in this work, at least in its current form.

\appendix

\section{Prerequisites}
\label{sec:prelim}
In this appendix we recall the basic tools of General Relativity (see e.g.\ \cite{Carroll:2004st}), to be applied in above sections. We use the Einstein index summation convention $a^\mu b_\mu \equiv \sum^3_{\mu=0}a^\mu b_\mu $ throughout.

Given a set of coordinates $x^\mu$ describing a space-time, along with the corresponding metric tensor $g_{\mu\nu}$ (lowering indices $x_\mu=g_{\mu\nu}x^\nu$) and inverse metric tensor $g^{\mu\nu}$ (raising indices $x^\mu=g^{\mu\nu}x_\nu$), the Christoffel symbols are determined as follows
\begin{align}
\label{Christoffel}
\Gamma^\sigma{}_{\mu\nu}=\frac{1}{2}g^{\sigma\beta}\left(\frac{\partial g_{\beta\mu}}{\partial x^\nu}+\frac{\partial g_{\beta\nu}}{\partial x^\mu}-\frac{\partial g_{\mu\nu}}{\partial x^\beta}\right).
\end{align}
The Ricci curvature tensor $R_{\mu\nu}$ and Ricci scalar $R$ are then constructed as
\begin{align}
R_{\mu\nu}=\frac{\partial\Gamma^\beta{}_{\mu\nu}}{\partial x^\beta} - \frac{\partial\Gamma^\beta{}_{\beta\nu}}{\partial x^\mu}+\Gamma^\beta{}_{\beta\alpha}\Gamma^\alpha{}_{\mu\nu}-\Gamma^\beta{}_{\mu\alpha}\Gamma^\alpha{}_{\beta\nu}~~~~~\text{and}~~~~~R=g^{\mu\nu}R_{\mu\nu}\,.
\end{align}
These quantities enter the Einstein field equations, along with energy momentum tensor $T^{\mu\nu}$ (a factor of $8\pi$ is absorbed into $T^{\mu\nu}$ here compared to other conventions)
\begin{align}
\label{EFE}
G^{\mu\nu}\equiv R^{\mu\nu}-\frac{R}{2}g^{\mu\nu}+\Lambda g^{\mu\nu} = \frac{G}{c^4} T^{\mu\nu},
\end{align}
where $G^{\mu\nu}$ is the Einstein tensor, and $\Lambda$ is a potentially non-zero cosmological constant. In the main text above, we work in geometrized units with the gravitational constant and the speed of light set to unity $G=1,\,c=1$.

For physical matter/radiation space-time content, the covariant divergence of the energy momentum tensor must vanish
\begin{align}
\label{covT}
0=\nabla_\nu T^{\mu\nu} = \frac{\partial T^{\mu\nu}}{\partial x^\nu}+\Gamma^\mu{}_{\sigma\nu}T^{\sigma\nu}+\Gamma^\nu{}_{\sigma\nu}T^{\mu\sigma}.
\end{align}

Free falling particles in a given space-time follow geodesic trajectories that are solutions to the geodesic equation
\begin{align}
\label{geodesic}
\frac{d^2 x^\mu}{d\tau^2}+\Gamma^\mu{}_{\alpha\beta}\frac{dx^\alpha}{d\tau}\frac{dx^\beta}{d\tau} = 0,
\end{align}
where $\tau$ is the proper time (or affine parameter) parameterizing the trajectory. Sometimes, it is also convenient to employ the proper time equation
\begin{align}
\label{propertime}
-d\tau^2=g_{\mu\nu}dx^\mu dx^\nu
\end{align}
to solve for a geodesic trajectory. 

\section*{Acknowledgments}
This work was supported by DOE grant DE-SC0011941.

\end{document}